\documentclass[preprint,preprintnumbers,amsmath,a4paper,aps,12pt]{revtex4-1}

\usepackage{bm}
\usepackage{graphicx} 
\usepackage{color}

\newcommand{\EEA}{\end{eqnarray}}              
\newcommand{\EEAN}{\end{eqnarray*}}

\begin{document}
 
\title{Effect of anisotropy on the glory structure of molecule-molecule 
scattering cross sections}

\author{Jes\'{u}s P\'{e}rez-R\'{i}os 
\footnote{Electronic mail: jpr@iff.csic.es}, Massimiliano Bartolomei,
Jos\'e Campos-Mart\'{\i}nez and Marta I. Hern\'andez}

\affiliation{Instituto de F\'{\i}sica Fundamental, 
Consejo Superior de Investigaciones Cient\'{\i}ficas (IFF-CSIC), Serrano 123, 
28006 Madrid, Spain}

\date{\today}

\begin{abstract}

Total (elastic + rotationally inelastic) integral cross sections are
 computed for O$_2(^3\Sigma_g^-)$-O$_2(^3\Sigma_g^-)$ using a recent ab initio
 potential energy surface. The sampled velocity range allows us a thorough
 comparison of the glory interference pattern observed in molecular
 beam experiments.  The computed 
cross sections are about 10\% smaller than the measured ones, however, a
 remarkable agreement in the velocity positions of the glory extrema is achieved. 
By comparing with models where the anisotropy of the interaction is reduced or
 removed, it is found that the glory pattern is very sensitive to the
 anisotropy, especially the positions of the glory extrema.

\end{abstract}

%\pacs{}

\maketitle

\section{Introduction}

   The velocity dependence of total integral cross sections in molecular beam
experiments usually shows an oscillatory pattern in the thermal and epithermal regimes. 
These glory undulations arise from a quantum-mechanical interference of
scattered waves for interaction potentials exhibiting wells, and indeed the
number of oscillations are related to the number of bound states of the
scattering potential\cite{Bernstein:62}. Glory undulations have been observed
in atom-atom\cite{Bernstein:73,Greene:72,Bernstein:67,Pritchard} and in  
atom-molecule\cite{Bernstein:61,Exp:68,ref1-Olson:68,o2kr} collisions, and
they have served to extract some features of the intermolecular interaction, in
particular, the spherically averaged or isotropic term of the
Potential Energy Surface (PES)\cite{Bernstein:73,Greene:72,Bernstein:67,o2kr,Agua}.  
Atom-molecule interactions do depend on the angles of orientation of the molecule
relative to the atom, and the effects of the anisotropy on
the cross sections, typically producing a damping of the glory
oscillations, have been widely studied
\cite{Exp:68,ref1-Olson:68,o2kr,Cross:68,Olson:68,Miller:69,Goldflam:78}.
However, and despite there have been
many experimental studies of molecule-molecule
collisions\cite{Perugia,Gomez:07,Cappelletti:08,Thibault:09}, the role of
anisotropy  for these more complex interactions
has been seldom studied from a theoretical point of view.

Recently, a high level global {\em ab initio} PES including the singlet,
triplet and  quintet 
multiplicities of the  O$_{2}\left(^{3}\Sigma_{g}^{-}\right)$
-O$_{2}\left(^{3}\Sigma_{g}^{-}\right)$ system has been reported by Bartolomei
{\em et al}\cite{JCP:2010}.  The interaction includes long-range coefficients 
obtained from first principles calculations of the electric properties of the
monomers\cite{J-Comp-Chem}. In order to study the performance of this PES, we
have carried out some dynamical calculations at different energy regimes
including ultracold\cite{jcp-ultracold:11}, cold\cite{Montero}, and
preliminary calculations in the thermal regime\cite{our-JPCA,JCP:2010}. For
cold translational temperatures, $10 \le T_{t}$ (K) $\le 34$, 
the quality of the {\em ab initio} PES was tested against measurements of
rotational populations of O$_2$ molecules traveling along a supersonic
jet\cite{Montero}. Close-coupling calculations
of rotationally inelastic state-to-state rate coefficients 
were found in good agreement with the experimental data,
 suggesting that the anisotropy of the {\em ab
  initio} PES is realistic. On the other hand, second virial
coefficients computed using the {\em ab initio} PES compare quite well
with experimental values over a wide range of
temperatures, and it was checked that including the full
anisotropy of the interaction is crucial to achieve such an
agreement\cite{JCP:2010,NIST}.  

In a higher energy regime,  Aquilanti {\em
  et al} reported total cross sections measurements for O$_{2}$-O$_{2}$
 using rotationally hot effusive beams\cite{Perugia}. These and 
analogous measurements using colder supersonic seeded beams, together with second virial
coefficient data, allowed the authors to obtain an experimentally derived
PES, the Perugia PES from now on.
In this way, cross section calculations just using the spherical average of the
Perugia PES showed quite a good comparison with the effusive beam
experiments\cite{Perugia}. However, analogous calculations using the recent
{\em ab   initio} PES showed some discrepancies with the experimental
data\cite{JCP:2010}. The effect of the anisotropy in the behavior of the total
integral cross sections has been only partially studied by means of
close-coupling calculations for a lower energy range and only using the
quintet PES\cite{our-JPCA}.

  In this work, we extend previous work\cite{JCP:2010,our-JPCA}
 by computing total (elastic + rotationally
inelastic) integral cross sections in the relevant energy range and
considering the fully anisotropic PES for the three spin multiplicities
involved. The coupled-states approximation\cite{McGuire-Kouri:74,Pack:74,Heil:78} is
used. Our aim is to provide a sensible test of the {\em ab initio}
PES in the thermal regime by comparison with the experiments of Ref.\cite{Perugia}.
In this regard, we also study how and how much the
anisotropy of the interaction affects the structure of the glory undulations
in the case of molecule-molecule collisions. 

 The paper is organized as follows.  In Section II, we give a short summary on the theory
and computational details. Results are reported and discussed in section III,
and conclusions are given in section IV.

\section{Theory and Computational details}

The collision dynamics of O$_2$+O$_2$ is studied within the
rigid rotor approximation and neglecting the fine structure. The theory for
the scattering of two identical linear rigid rotors has been reviewed in
Ref.\cite{our-JPCA} and applied to O$_2$+O$_2$
elsewhere\cite{our-JPCA,Montero}, so here we focus in some aspects
specific to this study. Using diatom-diatom Jacobi vectors
${\bf R}$, ${\bf r}_{1}$, and ${\bf r}_{2}$ in a space-fixed frame, the
Hamiltonian is written as (in atomic units) \cite{Green75,Miller}, 

\begin{equation}
H=-\frac{1}{2\mu R}\frac{\partial^{2}}{\partial R^{2}}R+
\frac{\hat{l}^{2}}{2\mu R^{2}}+
B_{e}\left(\hat{j}_{1}^{2}+ \hat{j}_{2}^{2}\right)+
V_M({\bf R},\hat{r}_{1},\hat{r}_{2}) 
\label{eq1}
\end{equation}

\noindent
where $B_{e}$= 1.438 cm$^{-1}$ is the rotational constant of the $^{16}$O$_{2}$ molecules, 
$\mu=15.9949$ amu is
the reduced mass of the collision system and $\hat{l}$, $\hat{j}_{1}$ and $\hat{j}_{2}$ are angular
momentum operators associated with  ${\hat R}$, $\hat{r}_{1}$ and
$\hat{r}_{2}$, respectively. 
Moreover, $V_M$ is the intermolecular PES for a given electronic spin of the 
complex ($M= S, T, Q$ for the singlet, triplet and quintet
spin multiplicities, respectively). The time-independent close-coupling
equations  are separately solved  for each spin
multiplicity (since the Hamiltonian commutes with the
total electronic spin) and also, for each symmetry block of the $G_{16}$ group, by using symmetry adapted
bases associated to the spatial inversion, permutation operator within the monomers and 
simultaneous permutation of nuclei between the monomers\cite{g16}. Statistical weights appropriate to  
$^{16}$O$_2$+$^{16}$O$_2$ are taken as explained in detail in
 Refs.\cite{our-JPCA,Montero}. 
Besides, for the present simulation of the number of particles lost from a molecular
beam due to scattering by a target, the computed integral cross section must
be corrected when the final states of the colliding partners are identical, in
order to avoid a double counting\cite{our-JPCA}.
In this way, the total (elastic + inelastic)
integral cross section for a selected pair of initial rotational states
($j_1,j_2$) is 

\begin{equation}
Q^{M}_{j_1,j_2} = \sum_{j_1',j_2'} \frac{Q^{M}_{j_1' j_2',
    j_1 j_2}}{1 +  \delta_{j_1',j_2'}},
\label{eq2}
\end{equation}

\noindent
where $Q^{M}_{j_1' j_2', j_1 j_2}$ is the state-to-state integral cross
section for indistinguishable monomers as obtained by
Takayanagi\cite{Takayanagi63}. These cross sections are finally averaged using
the multiplicity of each spin state as a statistical weight, 

\begin{equation}
Q_{j_1,j_2} = \frac{5 \, Q^{Q}_{j_1,j_2} + 3 \,
  Q^{T}_{j_1,j_2} + Q^{S}_{j_1,j_2}}{9}.
\label{eq3}
\end{equation}

Calculations have been carried out within the coupled-states (CS) 
 approximation\cite{McGuire-Kouri:74,Heil:78} as implemented in the {\sc molscat} 
code\cite{Molscat}.   
We used the global {\em ab initio} PES of Bartolomei {\em et
  al}\cite{JCP:2010} unless otherwise stated.
The cross sections of Eq.\ref{eq3} were computed for kinetic energies 
ranging from 27 to 4873 cm$^{-1}$. The close-coupling equations were solved
  using the Alexander and Monolopoulos's hybrid log-derivative/Airy 
propagator \cite{Hybridprop}. The propagation was carried out from 
 $R=$ 2.51 \AA $\,$ up to $R=46.8$ \AA, the switch between the log-derivative and 
the Airy propagator done at $R$= 11.7 \AA. Calculations were performed
 for increasing numbers of the total rotational angular momentum $J$ 
 until convergence (the highest $J$ reached was 606). For a given initial
 state,  $(j_1,j_2)$, a total of twelve rotational states were 
 included in the  expansion of the total wave
function, six of them having a internal energy lower than that of the initial
state. By performing additional CS calculations with four aditional rotational states,
 we have checked that the total cross sections are converged within about 2\%.

It remains to check the validity of the CS approximation. It is expected
that the CS approximation will work well when the kinetic energy becomes large
as compared with the potential well depth\cite{Kouri-cs,Heil:78}. We have compared
 the CS calculations with
 fully coupled calculations for kinetic energies ranging from 27 to
 360 cm$^{-1}$ (note that the O$_2$-O$_2$ well depth is of the order of 100
 cm$^{-1}$). The error obtained was smaller than $5\%$. For the higher energy
 range, where the fully coupled calculations would become extremely expensive,
 the error of the CS approximation should be smaller.

\section{Results and Discussion}

Present calculations are compared with the experimental results of Aquilanti
{\em et al}\cite{Perugia}. In that experiment, the oxygen projectile 
is in a hot effusive beam at a high rotational temperature (500 K), and it collides 
with an oxygen target in a reaction chamber at about 90 K. The most populated
rotational states in these conditions are $j_{1}$=9-13 and $j_{2}$=5 for beam
and target, respectively. Thus, we have chosen ($j_{1}$=11, $j_{2}$=5) as
the initial state for the cross sections of
Eq.\ref{eq3}. Some additional calculations were performed using other
populated states (e.g. $j_{1}$=9, $j_{2}$=7) and the results were
found to be insensitive to these modifications in the simulation.
The computed cross sections are obtained as functions of the kinetic
energy or, equivalently, of the velocity in the center-of-mass frame. In order
to compare with the experiments, they are transformed to functions
depending of the velocity in the laboratory frame according to the 
procedure described in Ref.\cite{Cappelletti:02}. 

\vspace{.5cm}

\begin{figure}[h]
\includegraphics[width=10.0cm,angle=0.]{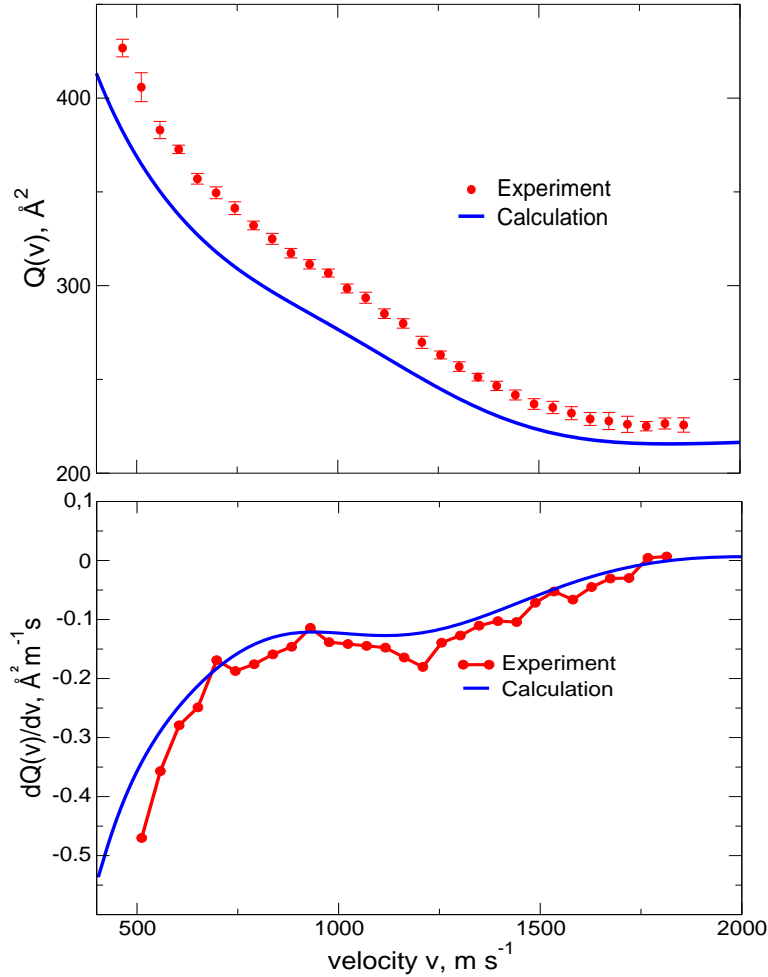}
\caption[]{ (Color online)
Upper panel: Total (elastic + inelastic) integral cross
  sections $Q(v)$ (in \AA$^{2}$) versus the velocity $v$ in the laboratory frame 
(in $m/s$). The experimental data \cite{Perugia} are given by circles, 
while present calculations using the {\em ab initio}
PES of Ref.\cite{JCP:2010} are shown using a solid line. 
Lower panel: Same as upper panel for the derivative of the total integral
 cross section with respect the velocity $dQ(v)/dv$ (in \AA$^{2} m^{-1} s$) as
 a function of $v$.} 
\label{fig1}
\end{figure}

Results are reported in Fig. 1 (upper panel), where computed and measured total
cross sections are plotted as functions of the velocity $v$ in the laboratory
frame.  It can be seen that the computed cross sections are
 smaller than the measured ones for the complete velocity range, the
 relative error varying from less than 15 \% for the lowest
velocities to about 5 \% for the highest ones. Although this could be termed
as a fair agreement, it should be noted that a closer agreement between {\em
  ab initio} calculations and experiment has been
achieved for other systems such as N$_2$-N$_2$, N$_2$-H$_2$ and
C$_2$H$_2$-H$_2$\cite{Gomez:07,Cappelletti:08,Thibault:09}. Possible reasons
for this relative discrepancy in the absolute cross sections are discussed below. 
On the other hand, it is readily noticed
that calculations and experiment agree quite well regarding the slopes 
of the curves as functions of the velocity. In the lower panel of
Fig.1 and 
in order to make a clearer comparison of the glory pattern, we show the
derivative of the cross sections with respect to the velocity. In this way,
the performance of the PES in the well region is more conveniently tested.
The agreement, it can be seen, can be termed as excellent. 
 We have found that the detailed interference pattern  is 
very sensitive to the anisotropy of the interaction. Indeed, agreement with
the observed positions of the undulations is
only achieved when the full anisotropy of the PES is included, as shown in
the following analysis. 

\vspace{0.5cm}

\begin{figure}[h]
\includegraphics[width=11.5cm,angle=0.]{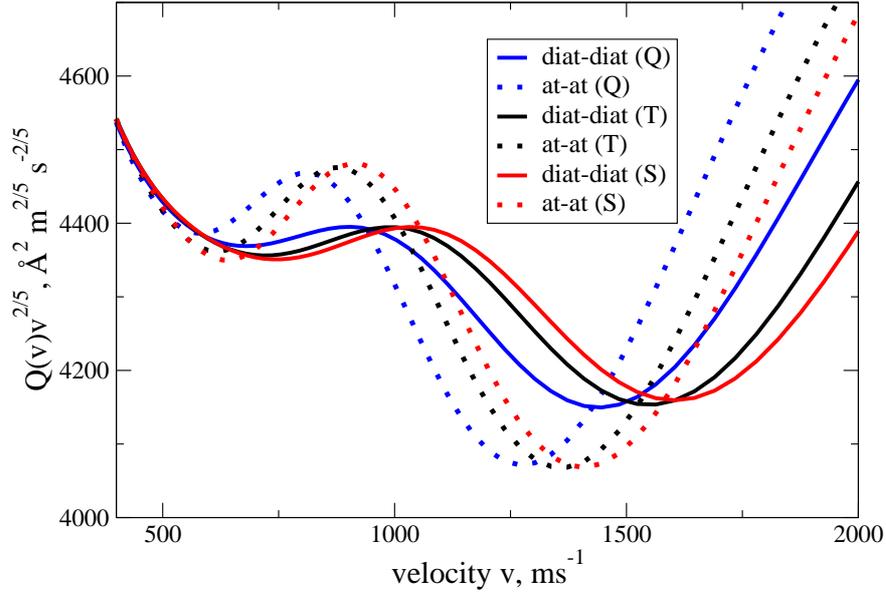}
\caption[]{(Color online) Total integral cross sections $Q(v)$
  multiplied by $v^{2/5}$  (in \AA$^{2}$ $(m/s)^{2/5}$) as functions of the
  laboratory velocity, $v$ (in $m/s$), for the different multiplicities of the
  {\em ab initio} PES\cite{JCP:2010} (Q, T, and S stand for quintet, triplet
  and singlet, respectively). Solid curves (``diat-diat'')  are present
  calculations where the full anisotropy of the PES is included, whereas
  dotted curves (``atom-atom'') refer to
 approximate calculations where only the spherically averaged interaction is
retained.} 
\label{fig2}
\end{figure} 

In Fig. 2 we present (solid lines) the individual cross sections for the 
singlet, triplet and quintet multiplicities, i.e., the ones entering in the
average of Eq. \ref{eq3}. The cross sections are now shown multiplied by
$v^{2/5}$  since it is
known\cite{schiff56,landau59} that the semiclassical $Q(v)\times v^{2/5}$ 
 is proportional to $C_6^{2/5}$ for  a $C_6/R^6$ interaction. In this way, the
glory structure is emphasized and the average value of $Q(v)\times v^{2/5}$
can be much easily related to the long range features of the
interaction\cite{Perugia}.  The calculations reported in Fig. 2 are
compared with an  
``atom-atom'' model (dotted lines), i.e., the cross sections for structureless
particles interacting with the spherically averaged potential of each of the
multiplicities. This approximation has been done by including only the 
 the $(j_{1},j_{2})=(0,0)$ rotational level and just the isotropic component of 
the interaction in the calculations. 

The analysis presented in Fig.2
 allows us to discuss the effects of both the {\em spatial} and the
{\em spin} anisotropy on the glory structure of the cross sections, so 
several comments follow. First, note that 
the average value (disregarding glory oscillations) of the {\em total}
cross section is very close to the average value of the (elastic) cross section in
absence of anisotropy, i.e., it essentially depends of the features of the
{\em isotropic} term of the interaction.  
This is the rule known as ``the conservation of total cross section'',
proposed and checked time ago\cite{Levine:72,Secrest:77}, and recently
confirmed for the present system by us\cite{our-JPCA}. 
We have additionally checked that such an average value agrees with an 
{\em effective} $C_6$ coefficient defined by the behavior of the isotropic
potential in the range $R=$ 6-8 \AA $\,$ (note that, in this region, singlet, triplet and
 quintet interactions are almost identical).  We have found that this
 ``effective'' $C_6$ value is due to contributions from $C_6$, $C_8$ and
 higher order long-range coefficients\cite{our-JPCA}.  
Given the sensitivity of the absolute cross sections to slight
modifications of the interaction energies, the fact that the
computed cross sections are somewhat smaller than the measured ones (Fig.1,
upper panel) could indicate some inaccuracies of the PES in that range of
 intermolecular distances.
%Further and more accurate {\em ab initio} calculations as well as experiments 
%should be devised for a better determination of the
%interaction energies.

A second aspect from Fig.2 is the modification of the glory structure due to
the anisotropy of the interaction. For a given spin multiplicity, the effect of the
anisotropy is a quenching of the oscillations as well as a
shift of the glory extrema, as compared with the isotropic, atom-atom model. 
The effect is found more significant as compared with previous  atom-diatom
studies\cite{Olson:68,Cross:68,Miller:69,Goldflam:78}, particularly in
the modification of the positions of the glory extrema (velocities giving
maxima or minima in the glory pattern). For instance, a similar analysis
for O$_2$-Kr concluded that the anisotropy reduces the glory amplitude but does
not modify such positions\cite{o2kr}. Finally and regarding the effect of the
 spin multiplicities, it is found that the positions of the glory
extrema vary with the different multiplicities (owing 
to differences in the well regions,
 the singlet and quintet potentials having the largest and smallest well depths,
respectively). This affects indeed the final glory structure after averaging
(Eq. \ref{eq3}). However, by comparing the effect of the anisotropy
 with that of averaging the different spin multiplicities,  
it can be concluded that the anisotropy plays a more determinant role in the
final shape of the cross sections.

 As has been shown, the addition of the anisotropy of the interaction does
  change the glory structure of the total cross sections and thus the
  atom-atom model is inappropriate for a realistic simulation of
  the experiments. This is a somewhat surprising conclusion in view of
  previous studies of effusive beams of diatomic
  molecules scattered by target atoms\cite{o2kr}. Diatoms emerge from a hot effusive
  beam with a high rotational temperature and therefore the rotation period
  becomes short as compared with the collision time. In these conditions, the
  effects of the anisotropy should be small and the interaction could be
  approximated by its spherical average. In addition, inelastic
  probabilities are expected to be small since the energy gap for a rotational
  transition becomes increasingly large as the initial rotational state
  increases.  The situation is different for diatom-diatom
  scattering since in this case the distribution of rotational levels becomes more
  uniform. For O$_{2}$-O$_{2}$, the average minimum energy gap is about
  20 cm$^{-1}$  for the relevant rotational levels, so that
  inelastic transitions may not seem such a rare event. 
  Moreover, analyzing the experimental conditions in more detail\cite{Perugia}, we find that
  the average rotational periods of beam and target are 
$\tau_{r}^{beam}\sim 1\times 10^{-12}$ $s$ and   
$\tau_{r}^{target}\sim 2\times 10^{-12}$ $s$, respectively.
The collision time ranges from $\sim 2 \times 10^{-12}$ $s$ 
at the lowest experimental velocity to $\sim 5 \times 10^{-13}$ $s$ at the
highest one. 
By comparing collision and rotational periods, we note that the 
O$_2$ in the beam may behave as a pseudo-atom for low velocities of the
beam. However, this approximation seems much more severe for the target
molecule, since its 
rotational period is larger and hence its relative orientation must play a role
in the collision dynamics, even at low velocities of the beam. 

\vspace{.75cm}

\begin{figure}[h]
\includegraphics[width=11.cm,angle=0.]{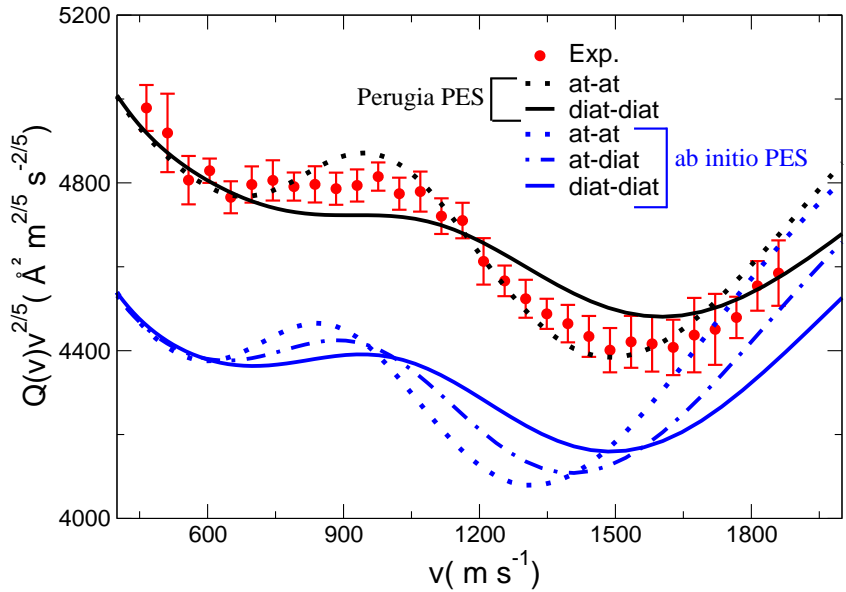}
\caption[]{(Color online) Total integral cross sections $Q(v)$ multiplied by 
$v^{2/5}$ (in \AA$^{2}$ $(m/s)^{2/5}$) as functions of the
  velocity ({\em ab initio} PES). The most accurate diatom-diatom calculation (full
anisotropy, in solid lines) is compared with the more approximate 
 atom-atom (isotropic) and  atom-diatom (partially
anisotropic) models (in dotted and dashed-dotted lines, respectively). 
 Points with error bars represent the data 
of Aquilanti {\em et al}\cite{Perugia}. Analogous calculations 
 calculations using the experimentally derived Perugia
PES\cite{Perugia} are also reported.}  
\label{fig3}
\end{figure}

To test these ideas, we have considered an atom-diatom collisional model, where the
atom represents the rotationally excited O$_2$ in the beam. To this end, we
have included a set of $(j_1=0,j_2=1-11)$ rotational levels in the
calculations, and have computed the total integral cross
section for the initial state $(j_1=0,j_2=5)$ using the CS approximation. 
In this way, the terms in the PES effectively involved in the simulation are equivalent
to those of an interaction potential between an atom and a rigid diatom.   
The total, spin averaged, cross sections are compared
with those of the isotropic (atom-atom) and the fully anisotropic
(diatom-diatom) models in Fig.3. 
Experimental data are also given in Fig. 3 as a guide to the eye.
As expected, the results of the atom-diatom model lie in between those of the
atom-atom and diatom-diatom calculations. While at low velocities of the beam all
the results are nearly the same, in the intermediate range (800-1200 $m/s$) 
the atom-atom model becomes poor but the atom-diatom
model still works well, indicating that in
this range the rapid rotation of the molecule in the effusive beam is
averaged. This approximation becomes worse for larger velocities ($v>$ 1200) where
the cross sections become more sensitive to the complete anisotropy of the
interaction. It is worth to note that the values of the velocity positions of glory
extrema vary very much with the different models studied, and that just the
most accurate calculation gives the best agreement with the experimental positions.
Finally, we have carried out calculations analogous to those explained here but
using the experimentally derived Perugia PES, and results for the atom-atom
model and the fully anisotropic calculation are also
displayed in Fig. 3. It can be seen that the atom-atom model agrees
quite well with the experiment: this is because some parameters of the
isotropic Perugia PES were indeed fitted to the observed total cross sections.
On the other hand, the effect of including the full anisotropy of the Perugia
PES is very similar to that already reported for the {\em ab initio} PES: the
glory amplitudes become quenched and positions of the extrema shift toward
higher velocities as compared with the atom-atom, isotropic model.

\vspace{.5cm}

\section{Conclusions}

We have carried out calculations of total integral cross sections for the
scattering of two rigid oxygen molecules aimed at testing a high level {\em ab
  initio} potential energy surface (PES) against effusive beam experiments in a
broad range of beam velocities. The calculations are in excellent agreement
with the measurements regarding the values of the beam velocities at which the
cross sections exhibit maximum or minimum values in the glory structure. 
We have shown that this agreement is
only achieved when the full anisotropy of the interaction is included in the
simulations. In this way and in contrast to accumulated experience in
atom-molecule scattering, it is reported that anisotropy in molecule-molecule
scattering not only produces a quenching in the glory amplitudes (with respect
to a purely isotropic interaction) but also a significant shift in the
positions of the glory extrema. 
This conclusion should be beared in mind when a potential inversion from
measurements is attempted in the case of molecule-molecule scattering.

As for a more global assessment of the O$_2$-O$_2$ {\em ab initio} PES, it can
be concluded that it is realistic as it reproduces very 
well the measured second virial coefficients, state-to-state rate coefficients
at low temperature, and the glory structure of the cross sections discussed here.   
This conclusion mainly refers to the short-range anisotropy of the interaction
 as this property is determinant for the observables
listed above. However the calculations underestimate (by about 10\%) the experimental
values of the total cross sections.  As mentioned in the previous section, it is the region
around 6-8 $\AA$ the responsible for the discrepancies between experimental and 
theoretical cross-sections.  In this region the dominant contribution corresponds
to the supermolecular calculation and, therefore, we would expect that a more
accurate ab-initio calculation (small changes of about tenths of cm$^{-1}$) will bring 
the theoretical results much closer to the experimental one.  In this regard it is worth 
noticing that among the three lowest surfaces corresponding to the singlet, triplet, and quintet 
multiplicities, only the latter has been computed at the 
highest level of theory, whereas the others have been obtained by differences
using a lower level of theory \cite{JCP:2010,PCCP-2008}. These complications would explain why
in similar cases it was possible a much closer agreement
between experiment and theory. Another possible factor 
might be the role played by the intramolecular spin-spin term (fine structure, neglected here) and 
its interplay with the different spin multiplicities of the intermolecular potential. Despite 
these possible improvements, it would be worthwhile to test
the present PES against more refined or new types of measurements.

%However, the calculations underestimate (by about 10\%) the
%observed absolute values of the cross sections. Then, a question may raise
%about the accuracy of the 6-8 \AA $\,$ range of the potential. Despite improved
%electronic structure and dynamics calculations should be attempted in the
%future, it would be worthwhile to test the present PES against either more refined or
%new types of measurements. 

\section{Acknowledgments}
  We acknowledge funding by Ministerio de Ciencia e Innovaci{\'o}n (Spain, 
  FIS2010-22064-C02-02). J.P.-R. is a JAE CSIC predoctoral fellow.

\end{document}